# Superspin Glass Aging Behavior in Textured and Non-Textured Frozen Ferrofluid

S. Nakamae[1,*], C. Crauste-Thibierge[1,¶], K. Komatsu[1], D. L'Hôte[1], Y. Tahri[1], E.Vincent[1], E. Dubois[2], V. Dupuis[2] and R. Perzynski[2]

[1]Service de Physique de l'Etat Condensé (CNRS URA 2464) DSM/IRAMIS, CEA Saclay
F-91191 Gif sur Yvette, France

[2]Physicochimie des Electrolytes, Colloïdes et Sciences Analytiques, UMR 7195, Université Pierre et Marie Curie
4 Place Jussieu, 75252 Paris, France

*Corresponding author: sawako.nakamae@cea.fr

¶Current address: Laboratoire Leon Brillouin (CNRS UMR 12) CEA Saclay, F-91191 Gif Sur Yvette, France

Keywords: superspin glass, ferrofluid, magnetic nanoparticles

ABSTRACT:

The effect of magnetic anisotropy-axis alignment of individual nanoparticles on the collective aging behavior in the superspin glass state of a frozen ferrofluid has been investigated. The ferrofluid studied here consists of maghemite nanoparticles ($\gamma$-$Fe_2O_3$, mean diameter = 8.6 nm) dispersed in glycerin at a volume fraction of ~15%. The low temperature aging behavior has been explored through 'zero-field cooled magnetization' (ZFCM) relaxation measurements using SQUID magnetometry. The ZFCM response functions were found to scale with effective age of the system in both textured and non-textured superspin glass states, but with markedly different scaling exponents, $\mu$. The value of $\mu$ was found to shift from ~0.9 in non-textured case to ~ 0.6 in the textured case, despite the identical cooling protocol used in both experiments.

## I. INTRODUCTION

It is now widely accepted that concentrated magnetic nanoparticles in solid media (*e.g.* frozen ferrofluid) can exhibit a transition from a superparamagnetic to a disordered collective state, called superspin glass (SSG).[1-3] The SSG state is believed to result from the frustration generated by dipole-dipole interactions among superspins (magnetic moments of nanoparticles) and from disorders in the system (*e.g.*, the random distributions of particles' positions, sizes and anisotropy-axis orientations). The 'flip-time' of individual superspins is much larger than that of atomic spins. Consequently, the accessible time scale for SSG dynamics is much shorter than that of atomic spin-glasses. Therefore, superspin glasses may very well be used to bridge the gap between the experimental time regime explored by the atomic spin glasses[4] and that by numerical simulations.[5] Moreover, concentrated frozen ferrofluids are valuable for understanding the SSG dynamics because one can selectively control physical parameters (*e.g.*, the interaction energy, the individual superspin size and the anisotropy alignment) and study their effects on the system. In magnetically textured frozen ferrofluids the magnetic easy-axes of all particles are uniformly fixed in space. Such a system suffers from one less disorder, namely, the distribution of anisotropy-axes. Only a few studies, both numerical and experimental, have been dedicated to the effect of texturing in concentrated magnetic nanoparticle assemblies at low temperatures.[6,7] These works show the existence of glassy metastable states, including SSG as well as other possible origins.

In the present paper, we study the anisotropy alignment effect on the out-of-equilibrium superspin glass dynamics of dense frozen ferrofluid in both textured and non-textured states by extracting the system's *effective age* from zero field cooled magnetization (ZFCM) relaxation measurements. In atomic spin glasses, the response functions obtained in the magnetization relaxation measurements are known to scale with waiting time, $t_w$, defined by the time elapsed from a temperature quench.[8] Similar scaling behavior has been found in frozen, non-textured ferrofluids.[9] As the anisotropy-axis orientation is the only difference between the textured and non-textured states studied, its impact on the SSG aging dynamics can be unambiguously identified by the direct comparison of their respective effective age as wall as their scaling behavior.

II. EXPERIMENT AND DATA ANALYSIS

The ferrofluids used in this study are made of maghemite, $\gamma$-$Fe_2O_3$, nanoparticles (~8.6 nm diameter)[10] dispersed in glycerin at ~15% volume fraction. Each nanoparticle is magnetically single-domain and bears an average permanent magnetic moment of ~$10^4 \mu_B$. More detailed sample descriptions are found elsewhere.[11] The ferrofluid was textured at 300 K and 3 T.[7] The existence of the SSG phase was determined from the frequency dependent ac susceptibility maxima that obey a critical law, characteristic of a SSG transition,[15] in both textured and non-textured states.[7,13]

To investigate the aging behavior, we carried out zero-field cooled magnetization (ZFCM) relaxation experiments (measured with a commercial SQUID magnetometer). In such experiments, samples are cooled from a temperature (140 K) that is well above the superspin-glass transition temperature, $T_g$ (~70 K for both SSG states), to the measuring temperature, $T_m$ (= 0.7 $T_g$ = 49 K) in zero applied field. After waiting for a period of $t_w$ (waiting time), a small magnetic field ($H$ = 0.5 Oe) is applied at $t$ = 0. The magnetization relaxation toward a final value, $M_{FC}$ (field cooled magnetization) is measured over a time, $t$, during which the relaxation rate also evolves, continuously changing the slope of the ZFCM response function. The relaxation rate, $S$, in a spin glass state is often expressed as the log-derivative of $M/M_{FC}$; i.e., $S = d(M/M_{FC})/d\log(t)$. The quantity $S$ is equivalent to the relaxation time distribution, a wide breadth of which is the reason behind the slow and non-exponential relaxation of the response function in a (super)spin glass state.[14,15] Generally, an inflection point in the ZFCM curve is produced at a characteristic time, $t_w^{eff}$, corresponding to the instance in time when the relaxation rate becomes the fastest, $S_{max}$. The position of $t_w^{eff}$ shifts depending on experimental control parameters such as $t_w$ and $H$, and therefore, it is commonly taken as an *effective age* of the system since the quench time. Furthermore, in superspin glasses, the normalized ZFCM can be separated into three independent components;[9] a stationary term, $m_{eq}(t)$, a superparamagnetic term, $m_{SPM}(t)$, and an aging term, $m_{ag}(t, t_w)$. The latter depends on both $t$ and $t_w$, and scales as a function of a reduced time variable of the form $\lambda/t_w^{\mu}$, where $\lambda = t_w[(1+t/t_w)^{1-\mu}-1]/[1-\mu]$ is an effective time which takes in account the $t_w$ dependent evolution of the magnetization relaxation.[8] Values of $\mu \neq 1$ indicate by how much the 'effective age' of a (super)spin glass deviates from the 'nominal age'; that is, experimental waiting time, $t_w$.

## III. RESULTS AND DISCUSSION

In Figure 1 (top panel), the inflection points, $t_w^{eff}$, observed in the ZFCM relaxation curves are plotted against the waiting time, $t_w$, on a *log-log* scale in both the non-textured and the textured SSG states. Both measurements were performed at $T_m = 0.7T_g$ with the excitation field $H = 0.5$ Oe and $t_w$ was varied between 3 and 24 ks. The generalized effect of aging in the magnetization relaxation is such that the longer $t_w$ is after a temperature quench, the slower the relaxation becomes. In atomic spin glasses, inflection points appear at $\log(t_w^{eff}) = \log(t_w)$ at low fields. As can be seen from the figure, $t_w^{eff}$ is indeed $\approx t_w$ in the non-textured SSG state. On the other hand, the values of $t_w^{eff}$ of the textured SSG state are significantly larger than the experimental $t_w$ for small $t_w$ values, indicating that the effective age is evolving differently from that of the experimentally recorded $t_w$. In fact, by adding an extra time, $t_{ini}$, to $t_w$; $t_w \rightarrow t_w + t_{ini}$, $t_{ini} \approx 1500$ s, the $t_w^{eff}$ plot of the textured SSG state coincides with that of the non-textured state (indicated by crosses). $t_{ini}$ may indicate that in the textured SSG, the *aging* had started during the cooling process, *i.e.*, ~1500 s prior to the experimentally defined quench time, $t_w = 0$. The cooling rate, however, was identical for both the textured and non-textured experiments. An analogy to the observed textured vs. non-textured 'effective age difference' may be drawn from the cooling-rate effect on atomic Heisenberg vs. Ising spin glasses. It has been reported that the effective age of an Ising spin glass *increased* after slower cooling, while Heisenberg spin glasses remained nearly insensitive to cooling-rate variations.[16] This analogy is particularly plausible as texturing causes the anisotropy-axis of all superpins to align, and therefore, the system should qualitatively approach an Ising-like magnetic state. This is consistent with the results obtained in our previous study[7] where the critical exponent, $z\nu$, associated with the critical law indicative of a SSG transition[12] was found to be slightly higher in the textured ferrofluid than in the non-textured case. Similarly, in atomic spin glasses, $z\nu$ values are larger in Ising spin glasses than in Heisenberg-like spin glasses.[4] It should also be noted that the prolonged effective age ($t_w^{eff} > t_w$) measured is not due to the applied field ($H = 0.5$ Oe) affecting the aging dynamics of the superspin glass. The magnetic field coupling can only *accelerate* the ZFCM relaxation process; resulting in the reduction of $t_w^{eff}$ rather than causing an increase as observed here.[13,17]

Another interesting influence of the anisotropy-axis alignment is found in the relaxation rate spectrum, $S$ (see Figure 1, bottom). The peak ($S_{max}$) width observed in the relaxation rate of the textured SSG state is much narrower than that of the non-textured case. As nanoparticles are magnetically single domained with a uni-axial anisotropy, the energy barrier distribution of a monodisperse nanoparticle assembly is expected to be more concentrated around a common value in a textured system, than in a non-textured system where anisotropy axis are oriented randomly. Consequently, the distribution of energy barriers of correlated superspin domains in the textured SSG state should also be narrower than the non-textured counter-part.

Figure 2 shows the ZFCM scaling of textured and non-textured under the measurement conditions described above. As mentioned earlier, the superparamagnetic ($m_{SPM}$) and the equilibrium ($m_{eq}$) components need to be subtracted in order to obtain a good scaling.[9] These contributions follow the forms of $B(\log(t/\tau_o^*))$ and $-A(t/\tau_o^*)^{-\alpha}$, respectively, where $B$ and $A$ are prefactors and $\alpha$ is a scaling exponent. In order to reduce the number of free parameters, we have fixed the value of $\tau_o^*$ ($\approx \tau_o \exp\{E_a/k_BT\}$, with $\tau_o$ is in the order of $10^{-9}$ s and $E_a$ is the anisotropy energy) to 5 µs.[13] The most important difference between the two scaling curves is in "$\mu$", the critical exponent appearing in the scaling variable $\lambda/t_w^\mu$ (see Introduction section). In atomic spin glasses; if $\mu = 1$ (that is, $t_w^{eff} = t_w$) then the system is said to be *fully aging*, if $\mu = 0$ then there is no aging (*i.e.*, $t_w$ has no effect on the magnetization relaxation of the system) and in-between values of $\mu$ reflect 'subaging'.[18] Although the exact mechanism behind subaging effects is not fully understood, it is known to be sensitive to cooling-rate history in the case of Ising spin glasses.[17] $\mu = 0.91$, found in the case of non-textured SSG agrees with the inflection point analysis where $t_w^{eff} \approx t_w$ was observed. This value is also close to the results obtained in a more concentrated maghemite ferrofluid[9] as well as to those found in atomic spin glasses.[8] In the textured SSG state studied, on the other hand, $\mu$ has been found to shift to a dramatically smaller value, 0.61.[19] This result confirms the smaller slope found in Figure 1 for the textured SSG state and it may also reflect, partly, the cooling rate effect as discussed previously in this section.

IV. SUMMARY

In summary, the zero field cooled magnetization relaxation behavior in a concentrated frozen ferrofluid was investigated in both magnetically textured and non-textured states. Anisotropy alignment was found to induce two effects; 1) subaging-like dynamics and 2) reduction in the spread of energy barrier distribution of correlated superspin domains. The scaling law, originally developed for atomic spin glasses and later modified to take in account the superparamagnetic particle contributions, was successfully applied in both cases but with a markedly different scaling exponent values. These results require further in-depth and more quantitative analysis which will be reported in a longer publication.

[19] Due to a multiple number of fitting parameters, slightly different solutions to $A$, $B$ and $\alpha$ can also produce reasonable scaling. However, $\mu$, which is the most influential on the overall scaling quality, must be near 0.9 (non-textured) and 0.6 (textured).

FIGURE CAPTIONS:

FIG 1: Top panel, $t_w$ vs. $t_w^{eff}$ of the ZFCM relaxation curves on a *log-log* scale. The cross symbols indicate the $t_w$ adjustments to $t_w \rightarrow t_w + t_{ini}$ where $t_{ini}$ = 1500 s (see text for details). Bottom panel, relaxation rate $S$ vs. $log(t)$ of the ZFCM curves of textured and non-textured ferrofluids at $H$ = 0.5 Oe and with $t_w$ = 3000 sec. The arrows indicate the positions of $S_{max}$(= inflection points). $M^* = M/M_{FC}$

FIG 2: (Color online) Scaling of ZFCM relaxation curves recorded at 0.7 $T_g$ in non-textured (top) and textured (bottom) SSG states for a series of waiting times, $t_w$. A superparamagnetic term stemming from a finite particle size distribution [$B\log(t/\tau_o^*)$] and an equilibrium contribution from smaller nanoparticles [$-A(t/\tau_o^*)^{-\alpha}$] are subtracted from the total ZFCM. The scaling procedures are explained in text.

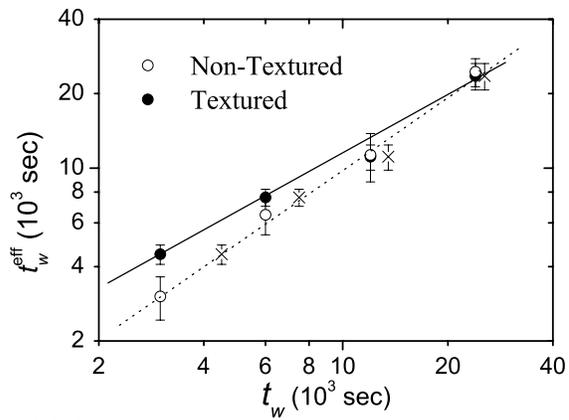

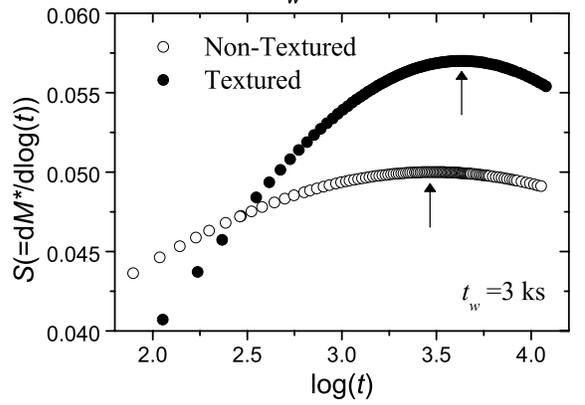

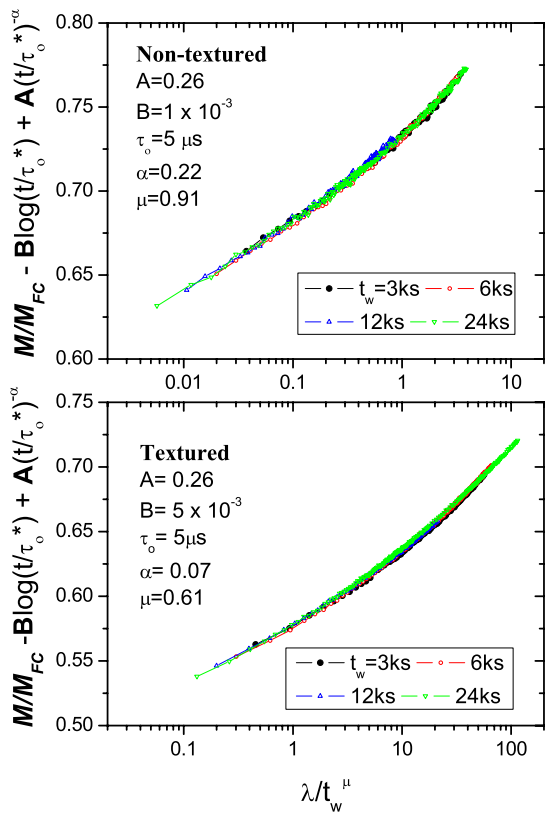